\documentclass[twocolumn,prl,showpacs,floatfix]{revtex4}

\usepackage{graphicx}

\begin{document}

\title
{
Discreteness-Induced Criticality in Random Catalytic Reaction Networks 
}
    
\author
{ 
Akinori Awazu$^1$ and Kunihiko Kaneko$^{2}$
}

\affiliation
{
$^1$Department of Mathematical and Life Sciences, Hiroshima University,
Kagami-yama 1-3-1, Higashi-Hiroshima 739-8526, Japan.
}

\affiliation
{
$^2$Department of Basic Science, University of Tokyo \& ERATO Complex 
Systems Biology, JST, 
Komaba, Meguro-ku, Tokyo 153-8902, Japan.
}

\date{\today}

\begin{abstract}
Universal intermittent dynamics in a random catalytic reaction network,
induced by smallness in the molecule number is reported. 
Stochastic simulations for a random catalytic reaction network subject 
to a flow of chemicals show that the system undergoes a transition from 
a stationary to an intermittent reaction phase when the flow rate is 
decreased.
In the intermittent reaction phase, two temporal regimes with active and 
halted reactions alternate.
The number frequency of reaction events at each active regime
and its duration time are shown to obey a universal power laws with
the exponents
$4/3$ and $3/2$, respectively. These power laws are explained
by a one-dimensional random walk representation of the number of 
catalytically active chemicals. Possible relevance of the result to 
intra-cellular reaction dynamics is also discussed.
\end{abstract}


\maketitle

Most intra-cellular reactions progress with the aid of catalysts. These 
reactions are essential for a cell to survive and grow. All catalysts 
that are proteins are synthesized within a cell as a result of catalytic 
reactions. 
Thus, researchers have started investigating catalytic reaction networks in 
order to develop a theoretical model of intra-cellular dynamics and 
protocells for their role in the origin of life\cite{Eigen,Kauffman,Stadler,Lancet,Jain,Furusawa,Ito,Kaneko-Adv,Kaffuman_PRE}; research in this field was 
pioneered by Kauffman\cite{Kauffman}.  
Recent studies on a growing cell model comprising random catalytic reaction 
networks have also revealed universal statistical laws on chemical abundances.
These results are confirmed by the gene expression
data of the present cells\cite{Furusawa,Ito}. 

Generally, cells consist of a large number of chemical species,  
some of which flow in and out through the membrane. There are also other 
chemical species that are not present in a large quantity.
However, some chemical species play an important role even at extremely 
low concentrations amounting to only a few molecules per 
cell\cite{H_M,cell2,cell3,cell4}. 
In such cases, the fluctuation and discreteness in the molecule number 
are very important as far as temporal variations and steady distribution 
of chemical concentrations are considered. Indeed, recent studies on 
catalytic reaction dynamics have demonstrated that
discreteness in molecule number can induce drastic changes in the 
temporal variations and the steady distribution of chemical concentrations 
obtained from the rate equation model\cite{Shnerb,togashi1,ookubo,DIT}.
However, in most thermodynamics studies, the molecule number is assumed 
to be large while the number of species is assumed to be rather small.
In order to study the catalytic reaction network of a cell, it is important 
to consider the case a large number of chemical species and a small molecule 
number.

In this study, we investigate a simple model of a random catalytic reaction 
network subjected to a flow of chemicals. In a region with a weak flow, 
where the discreteness in molecule number is large,
it was observed that the reaction events occurred intermittently, separated 
by a quiescent state.  It was also observed that the number distribution 
of reaction events during each reaction state obeyed a universal 
power law, displaying a feature of criticality induced by the discreteness 
in the molecule number.  By referring to the study of self-organized 
criticality, as investigated in models for sand-piles, earthquakes, 
interface depinning in random media, and so 
on\cite{SOC1,SOC2,SOC3,SOC5,SOC6,SOC-Jensen}, we will explain the origin 
of this universal power law behavior. Finally, we discuss in brief the 
possible relevance of our result to the power laws of the distribution of 
residence time in the quasi-stationary state observed in behaviors of cells.

\begin{figure}
\begin{center}
\includegraphics[width=7.0cm]{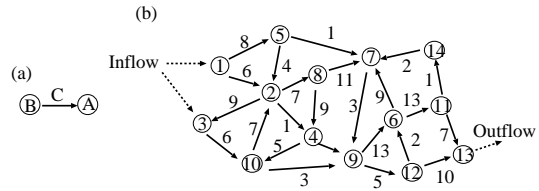}
\end{center}
\caption{(a) An illustration of a catalytic reaction $B+C \to A+C$, and (b)
an example of a catalytic reaction network.}
\end{figure}

We introduce a network of elementary, two-body 
catalytic reactions among a large number of chemical species\cite{Furusawa}. 
Assuming that chemicals in the system are well stirred, we discard the 
spatial dependence of concentration so that the state of the system can be 
represented by a set of numbers $(n_1, n_2, .... n_M)$, where $n_i(0,1,...)$ 
indicates the number of molecules of each chemical species 
$i$ ($1 \le i \le M$), with $M$ being the total number of chemical species. 
Each chemical is transformed to another by (several) reactions. These 
reactions are catalyzed by another chemical, i.e., the reaction from 
chemical $B$ to chemical $A$ is catalyzed by a third chemical $C$
(see Fig. 1).  For simplicity, the reaction rates $r$ of each 
reaction are set to be identical (the inhomogeneous reaction rate case is 
discarded as it does not affect the result to be discussed). 
The growth rate of molecule number $n_A$ (or the decay rate 
of molecule number $n_B$) is given by $r n_B n_C / V$, on 
the average, where $V$ represents the volume of the system. 

The entire catalytic reaction network consists of reactions whose 
reaction paths are chosen randomly (and then fixed). The average 
number of reaction paths from a chemical $i$ catalyzed by a chemical 
$j$ is set to a given connection number $K$. In this study, we only 
consider the case where $K_c < K << M$; $K_c$ indicates the 
critical connection number at the percolation threshold in random 
networks ($K_c\sim \log(M)$) \cite{grahu}.  
We have not considered an auto-catalytic reaction of the form 
$B + C \to 2C$ because usually such a reaction is not elementary and 
is realized as a result of a series of (non-auto-catalytic) elementary 
reactions. 

Next, we consider the flow of chemicals into a system from a molecular 
bath. We assume that $M_{in}$ chemical species can flow into the 
system. At each instance, a single molecule from one of the $M_{in}$ 
chemical species is added to the system in the rate $Q/M_{in}$, where 
$Q$ indicates the inflow rate of molecules. For simplicity, the flow 
rates of all chemical species are assumed to be the same; however, this 
assumption can be relaxed. We also assume that 
$M_{out}$ chemical species flow out (or are decomposed) from the 
system. 
Because of this inflow and outflow of chemical species, the total number 
of molecules $N=\sum_j n_j $ varies with time, which is in contrast to 
our previous study\cite{DIT}.

In this study, we assume that $M_{in}$ is sufficiently larger than 
$\sqrt{M/K}$. It should be noted that the average number of reaction paths 
through which the $M_{in}$ 'input' chemicals are catalyzed by one of 
themselves is given as $~ M_{in}K \times M_{in}/M$.
This number is much greater than unity when $M_{in} >> \sqrt{M/K}$. 
As long as this condition is satisfied (empirically, 
$M_{in} \stackrel{>}{\sim} 2\sqrt{M/K}$), a chemical reaction takes place 
even if the total molecule number is null initially. With this
constraint on $M_{in}$, the behavior of the system is observed independent 
of $M_{in}$. 
Here, we mainly show the results of the system behavior for $M_{in} = M$.
Our second assumption is that $M_{out}$ is much smaller than $M$. If the 
value of $M_{out}$ is comparable to that of $M$, the number of extinct 
chemical species increases. As a result, the number of surviving chemical 
species is much smaller than $M$, thereby reducing the effective system size 
considerably. 
On the other hand, if this assumption is true, the qualitative behavior 
of the system is determined by $M$. 
In this study, we mainly present the results of the system behavior for 
$M_{out} = 1$.

We perform stochastic particle simulation in order to study the possible 
effects of stochasticity and discreteness in molecule numbers. The simulation 
procedure is as follows: At each simulation step, we randomly select a pair 
of molecules. If the pair consists of a substrate and one of its catalysts, 
then, according to the rule of the catalytic reaction network, the substrate 
molecule is replaced by the product molecule with the probability $r$. The 
molecule number $n_i$ of input chemical ($1 \le i \le M_{in}$) is incremented 
by one with a probability of $Q/M_{in}$ per unit time. When the output 
chemical species $j$ ($(M-M_{out}) < j \le M$) is generated by the 
reaction or the inflow of molecules, it is removed immediately. 
We have adopted Gillespie's direct method in order to carry out numerical 
simulations for this process\cite{Gillespie}.

\begin{figure}
\begin{center}
\includegraphics[width=8.0cm]{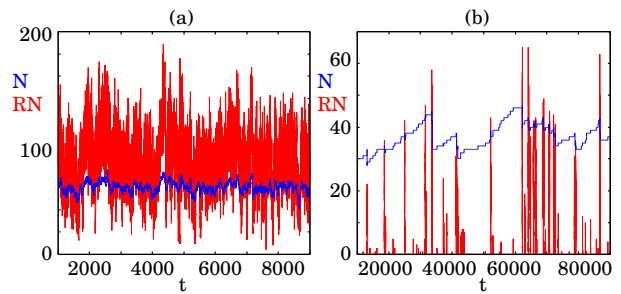}
\end{center}
\caption{(Color online) Typical temporal evolutions of $N(t)$ 
(blue) and $RN(t)$ (red) for $M=300$ and $K = 12$ with (a) $Q=0.3$ 
and (b) $Q=0.001$.} 
\end{figure}

Here, we present the results of numerical simulations obtained from
a variety of random networks and several sets of parameters. 
Without losing generality
(by redefining the time unit), we can assume $r = 1$ and $V = 1$.
Figure 2 shows the plots of typical temporal evolutions of the total number 
of molecules in the system, $N(t)$, and the reaction number, 
$RN(t)$, for $t = 0, 1, 2,...$ for $M = 300$, and $K = 12$ with 
(a) $Q = 0.3$ and (b) $Q=0.001$. Here, $RN(t)$ is defined as the number 
of reaction events that have occurred between $t$ and $t + 1$. When $Q$ 
is large, $N$ and $RN$ show stationary, relatively small fluctuation 
around their mean values. On the other hand, when $Q$ is small, there 
are intermittent bursts in $RN$, while $N$ increases gradually, until the 
increase is replaced by the drastic drop in $N$ as a 
result of such bursting reactions. These two regimes alternate.

The reason for change in behaviors of $N(t)$ and $RN(t)$ because of the 
decrease in $Q$ is as follows. When $N$ is not large enough due to low 
inflow rate $Q$, the molecules therein do not react with 
each other because catalyst molecules for each of existing chemical species 
are absent. The reaction then stops and $RN$ is constant at 
$0$, until a flow of necessary catalysts restarts the reaction process.
When the reactions start, the process of production and elimination of 
catalysts is repeated until the catalysts for all molecules in the system 
disappear. 

If $Q$ is sufficiently large, the system is in a steady state expected 
from the fixed point solution of the continuous rate equation. There exists 
a deviation in $N$ and $RN$ by stochastic inflows and the reaction process 
to produce the chemicals that escape from the system. However, these 
deviations are not substantial. Thus, a steady state with large $N$ is 
sustained so that the reaction does not stop.

\begin{figure}
\begin{center}
\includegraphics[width=8cm]{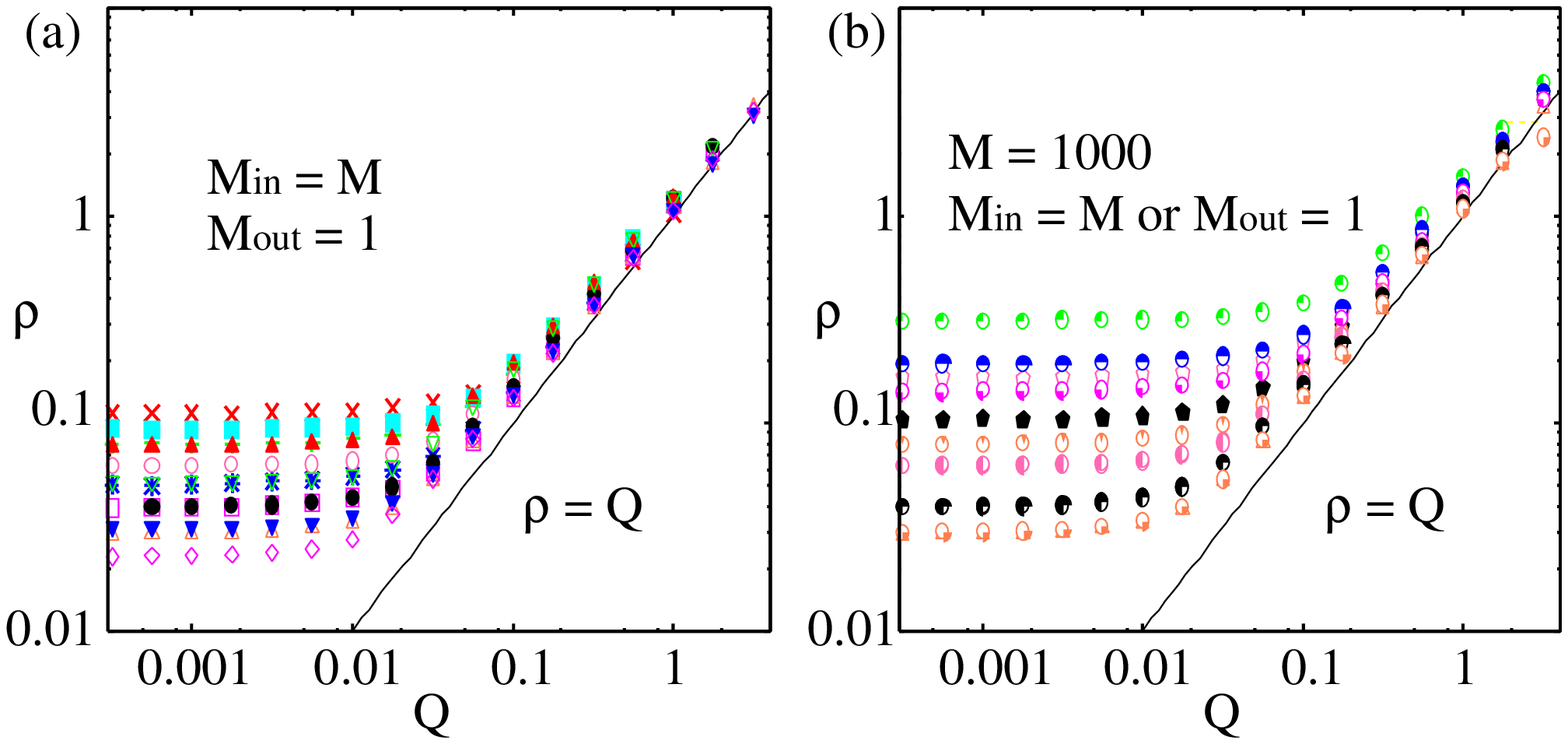}\\
\includegraphics[width=8.0cm]{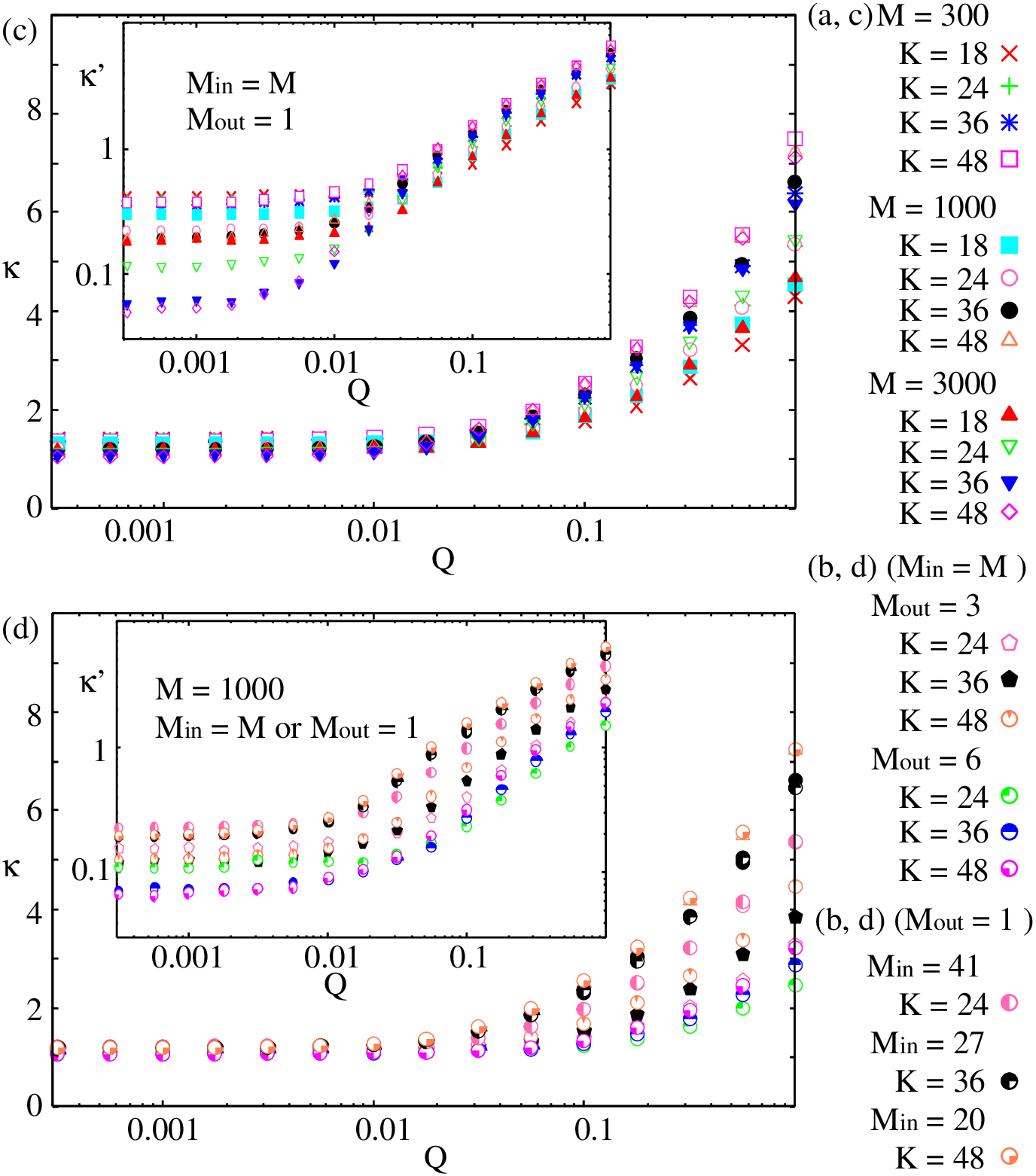}
\end{center}
\caption{(Color online) $\rho$, $\kappa$, and $\kappa'$ as a 
function of $Q$ for several $M$, $K$, $M_{in}$, and $M_{out}$. $\rho$, 
$\kappa$, and $\kappa'$ are computed by sampling with a step size of 
$10^7$.} 
\end{figure}

\begin{figure}
\begin{center}
\includegraphics[width=8.0cm]{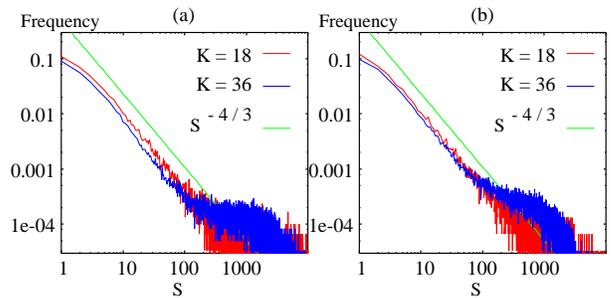}
\end{center}
\caption{(Color online) Frequencies of $S$ for $M = 1000$, $K = 18$ and 
$36$, $Q=0.0003$ with (a) $M_{in} = M$ and $M_{out} = 1$ and (b) 
$M_{in} \sim M/K$ and $M_{out} = 3$ ($M_{in} = 55$ for $K = 18$ and 
$M_{in} = 27$ for $K = 36$). 
These histograms are computed by sampling with a step size of $10^8$. }
\end{figure}

We have plotted ${\bar N}$, the long time average of $N$ versus the change 
in $Q$ for statistical characterization of the transition mentioned above. 
Figs. 3 (a) and (b) are the log-log plots of $\rho = {\bar N}^{2}KM_{out}/M^{2}$ 
versus $Q$, whereas Figs. 3 (c) and (d) are the plots of 
$\kappa = \frac{{\bar N}K}{M}$ and $\kappa' = \kappa -1$ as functions of 
$Q$ for several $M$, $K$, 
$M_{in} = M$ and $\sim M/K$, and $M_{out} = 1$, $3$, and $6$. 
Here, Figs. 3(a) and (c) show the results for $M_{in} = M$ and $M_{out} = 1$ 
and Figs. 3(b) and (d) show the results under $M = 1000$. 

In Figs. 3(a) and (b), $\rho \approx Q$ holds for large 
$Q \stackrel{>}{\sim} M_{out}/K$. This can be explained as follows. 
In the steady state, 
the average molecule number for each chemical is ${\bar N}/M$, which 
also gives the number of catalysts per reaction path. Then, the 
rate of production of each chemical is $\sim K (  {\bar N}/M)^2$. Thus the 
outflow rate is estimated as $M_{out} \times K( {\bar N}/M)^2= \rho$ which 
should balance with the inflow rate $Q$ at the steady state.

On the other hand, for a small $Q$, ${\bar N}$ is constant 
independent of $Q$. This is explained by the estimate of the number of 
molecules for the discreteness-induced transition\cite{DIT}.  Here, the 
reactions tend to stop if the average number of reaction paths that have 
corresponding non-zero catalyst for each chemical is smaller than 
unity, i.e., if $\frac{NK}{M} \stackrel{<}{\sim} 1$.
In this case, the outflow of molecules is suppressed; however, $N$ 
increases because of the inflow of molecules, until it satisfies the 
condition $\frac{NK}{M} \stackrel{>}{\sim} 1$. Hence, $\bar{N}$ remains at 
a constant value that is slightly larger than $ M/K$ for small $Q$. 
Indeed, Figs. 3(c) and (d) for small $Q$ indicate that the $Q$ versus 
$\kappa$ and $Q$ versus $\kappa '$ relations for a large $K$ are fitted by 
a single curve for each $M$ and $M_{out}$, independent of $K$ and $M_{in}$,
while $\kappa'$ approaches $0$ with an increase in $M$ and $M_{out}$. 
Moreover, the crossover point between the above two states is estimated as 
$Q_c \sim M_{out}/K$.

We now focus on the statistical properties of the system at the regime of 
small $Q$ with a constant ${\bar N}$. In this regime, the inflow of molecules 
continues even after the successive reactions have been completed. This 
situation will continue until an inflow triggers the next successive 
reactions, as shown in Fig. 2(b). Then, we focus 
on statistical properties of the reaction size $S$. $S$ is defined as the 
sum of $RN(t)$ between the interval of quiescent states where the molecules 
therein can not react with each other.
Figure 4 shows the frequency of the reaction size $S$, for a system with
$M = 1000$, $K = 18$ and $36$, $Q=0.0003$, with $M_{in} = M$ 
and $M_{out} = 1$ for (a) and $M_{in} \sim M/K$ and $M_{out} = 3$ for (b). 
As shown, the frequency distribution of $S$, $P(S)$, obeys a universal 
power law $P(S) \sim S^{-\gamma}$ with the exponent $\gamma \sim 4/3$ for 
small enough $Q$. The same universal power law is satisfied even for other 
values of $M$, $M_{in}$, $M_{out}$, and $K$.

Now we discuss the origin of this universal power law for a small $Q$. 
Let us split the chemical species into active and inactive ones. 
An active chemical species is defined as that species which is present in 
the system along with its catalysts.
When $M_A(t)=0$, the system is said to be in the inactive state, and when 
$M_A(t)>0$, the system is said to be in the active state.

During the inactive state, $M_A(t)$ increase
due to the inflow of some molecules. Reactions start taking place due to 
this inflow and production and elimination of several 
chemicals (catalysts) are repeated leading to an increase and then a 
decrease in $M_A(t)$. This process continues until $M_A(t)$ returns to 0, 
i.e., when the catalysts for all the molecules 
in the system have disappeared.  
Let us define the duration of reactive state $D$
as the time interval with $M_A(t) > 0$. Note that $RN(t)$ is considered to be 
proportional to $M_A(t)$ on average. 

Since the system under consideration is a random catalytic reaction network, 
we assume that the change of $M_A$ is approximated by a one-dimensional 
random walk with $0 \le M_A \le M$. Then, the frequency of $D$ is given by 
$P(D) \sim D^{-\alpha} = D^{-3/2}$, on the basis of the first return time 
distribution for the one-dimensional random walk. The reaction size $S$, 
which is proportional to the area below the random walk before its
first return, scales as $S \sim D^{\beta} = D^{3/2}$. 
Then, the exponent $\gamma$ for $P(S) \sim S^{-\gamma}$, 
is given by the standard relation between the 
exponents $\alpha$, $\beta$, and $\gamma$: 
$\gamma = 1+(\alpha-1)/\beta = 4/3$\cite{SOC5}. 

\begin{figure}
\begin{center}
\includegraphics[width=8.0cm]{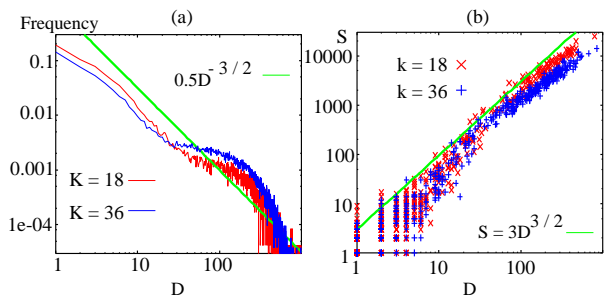}
\end{center}
\caption{(Color online) 
(a) Frequencies of $D$ for $M=1000$, $Q = 0.0003$, $M_{in} = M$, and 
$M_{out} = 1$ with $K=18$ and $K=36$ and (b) $D-S$ characteristics for the same 
conditions. The histogram (a) is computed by the sampling with a step size of 
over $10^8$ and (b) is computed with a step size of $2 \times 10^6$.}
\end{figure}

In order to confirm the validity of the above argument, we 
have examined the frequency of $D$ and the relationship between $D$ and 
$S$. Figure 5 (a) shows the frequency of $D$ and Fig. 5 (b) shows the $D-S$ 
characteristics for 
$K=18$ and $K=36$ with $M=1000$, $M_{in} = M$, $M_{out}=1$, and $Q=0.0003$, 
which are obtained from the same simulation as 
the data plotted in Fig. 4(a). The Figs. 5 (a) and (b) support the relation 
$P(D) \sim D^{-3/2}$ and $S \sim D^{3/2}$. Note that a similar random-walk 
description was adopted in a model of  self-organized criticality in the 
anisotropic interface depinning in a quenched random medium\cite{SOC5}. 

In this study, the dynamic aspects of a random catalytic reaction network 
subjected to a flow of chemicals was studied. With a decrease 
in the inflow rate, the system undergoes a transition from a stationary 
to an intermittent reaction state; however, discreteness in molecule number 
in the latter state is essential. The frequency of the reaction size and 
duration of the reactive states obey the universal power laws with 
exponents $4/3$ and $3/2$, respectively. Note that this critical 
behavior is obtained without any tuning parameter values, and as long as 
the inflow rate is small enough to maintain the discreteness (0,1..) in 
the molecule number.
In other words, discreteness induced self-organized criticality 
(SOC)\cite{SOC1,SOC2,SOC3,SOC5,SOC6}.

Power law of the residence time distributions in a certain state are 
often reported in the behavior of cells, such as the residence time at 
a counter-clockwise rotations of flagella motor in E-coli\cite{e-coli}, 
the frequency of the inter-event intervals for spontaneous vesicular 
release at a Xenopus neuromuscular junction\cite{neuron}, and the frequency 
of the inter-beat time interval in cardiac muscle cell\cite{harada}, where
non-stationary behaviors are a result of intra-cellular reaction dynamics.
Since our random reaction network model is rather simple, we cannot make a 
detailed comparison with these experiments; however, the discovery of a 
universal power law may help us to understand the ubiquity of non-stationary 
intra-cellular 
dynamics and power law behaviors.

This research of A. A. was supported in part by a Grant-in 
Aid for Young Scientist (B) (Grant No. 19740260).

\end{document}